\documentstyle[12pt]{article}
\begin{document}
\tolerance=5000
\def\be{\begin{equation}}
\def\ee{\end{equation}}
\def\bea{\begin{eqnarray}}
\def\eea{\end{eqnarray}}
\def\nn{\nonumber \\}
\def\cF{{\cal F}}
\def\det{{\rm det\,}}
\def\Tr{{\rm Tr\,}}
\def\e{{\rm e}}
\def\etal{{\it et al.}}
\def\erp2{{\rm e}^{2\rho}}
\def\erm2{{\rm e}^{-2\rho}}
\def\er4{{\rm e}^{4\rho}}
\def\etal{{\it et al.}}

\  \hfill
\begin{minipage}{2.5cm}
NDA-FP-48 \\
May 1998 \\
\end{minipage}

\

\vfill

\begin{center}

{\large\bf CAN QUANTUM-CORRECTED BTZ BLACK HOLE ANTI-EVAPORATE?}

\vfill

{\large\sc Shin'ichi NOJIRI}\footnote{
e-mail : nojiri@cc.nda.ac.jp}
and
{\large\sc Sergei D. ODINTSOV$^{\spadesuit}$}\footnote{
e-mail :
odintsov@tspi.tomsk.su}
\vfill

{\large\sl Department of Mathematics and Physics \\
National Defence Academy \\
Hashirimizu Yokosuka 239, JAPAN}

{\large\sl $\spadesuit$
Tomsk Pedagogical University \\
634041 Tomsk, RUSSIA \\
}

\vfill

{\bf ABSTRACT}

\end{center}

Kaluza-Klein reduction of 3D gravity with minimal scalars 
leads to 2D dilaton-Maxwell gravity with dilaton coupled scalars. 
Evaluating the one-loop effective action for dilaton coupled scalars 
in large $N$ and $s$-wave approximation we apply it to study quantum 
evolution of BTZ black hole. It is shown that quantum-corrected BTZ BH 
may evaporate or else anti-evaporate similarly to 4D Nariai BH 
as is observed by Bousso and Hawking. Instable higher modes in the 
spectrum indicate also the possibility of proliferation of
BTZ BH.

\ 


\newpage

\noindent
1. BTZ black hole (BH)\cite{BTZ} attracts a lot of attention due to various 
reasons. In particulary, it is related via T-duality with a class 
of asymptotically flat black strings \cite{H} and via U-duality 
it is related \cite{SH} with 4D and 5D stringy BHs \cite{HS} 
which are asymptotically flat ones. Hence, microscopically computing 
the entropy of BTZ BH \cite{SK} may be applied via duality relations 
for the computing of entropy for higher dimensional BHs. 
This fact is quite remarkable as above types of BHs look completely 
different from topological, dimensional or space-time points of view. 

BTZ BH being locally ${\rm AdS}^3$ without curvature singularity 
may be considered 
as a prototype for general CFT/AdS correspondence \cite{MW}. Indeed,
 3D gravity has no local dynamics but BH horizon induces an effective 
boundary (actually, 2D WZWN model). Hence,quantum studies around BTZ BH 
may help in better understanding of above correspondence.

In the present letter we study quantum dynamics of BTZ BH 
due to 3D minimal scalars. We work in large $N$ and $s$-wave approach
 where minimal scalars are described as 2D dilaton coupled 
scalars. The quantum corrected version of BTZ BH is found.
It is shown that it can evaporate or anti-evaporate 
in accordance with recent observation by Bousso and Hawking 
\cite{BH2} made for 4D SdS BH. Higher modes 
perturbations in the quantum spectrum are also briefly discussed.

\ 

\noindent
2. Three dimensional Einstein gravity with the cosmological term 
has the exact rotating black hole solution \cite{BTZ} and the solution 
can be regarded as an exact solution of string theory \cite{H}.
In this section, we consider the circle reduction of 
three dimensional Einstein gravity to two dimensional one.
Now we identify the coordinates as $x^1=t$, $x^2=r$, $x^3=\phi$.
Here $t$, $r$ and $\phi$ are the coordinates of
time, radius, and angle.
We now assume that all the fields do not depend on $x^3$.
Then, under the following infinitesimal 
coordinate transformation $x^3\rightarrow x^3 + \epsilon (x^1,x^2)$,
the metric tensors transform as follows;
\be
\label{metrictrnsf}
\delta g^{(3)3\alpha}=\delta g^{(3)\alpha 3}=
g^{(3)\alpha\beta}\partial_\beta\epsilon\ ,\ \ 
\delta g^{(3)33}=2g^{(3)3\alpha}\partial_\alpha\epsilon\ ,\ \ 
\delta g^{(3)\alpha\beta}=0
\ee
Here $\alpha, \beta=1,2$. 
Eq.(\ref{metrictrnsf}) tells that we can identify the 
gauge vector field $A_\alpha$ with
$g^{(3)3\alpha}=g^{(3)\alpha\beta}A_\beta$.
Eq.(\ref{metrictrnsf}) also tells that the metric 
tensor $g^{(3)\mu\nu}$ ($\mu,\nu=1,2,3$) 
can be parametrized a la Kaluza-Klein as
\be
\label{3dmtrc}
g^{(3)\mu\nu}=\left(
\begin{array}{cc}
g^{\alpha\beta} & g^{\alpha\gamma}A_\gamma \\
g^{\beta\gamma}A_\gamma & \e^{2\phi}
+g^{\gamma\delta}A_\gamma A_\delta \\
\end{array}\right)\ .
\ee
Then we find 
\bea
\label{g3}
g^{(3)}&\equiv&{1 \over \det g^{(3)\mu\nu}}=g\e^{-2\phi} \ \ 
(g\equiv {1 \over \det g^{\mu\nu}}) \\
\label{ginv}
g^{(3)}_{\mu\nu}&=&\left(
\begin{array}{cc}
g_{\alpha\beta}+\e^{-2\phi}A_\mu A_\nu & -A_\gamma \\
-A_\gamma & \e^{-2\phi} \\
\end{array}\right)\ .
\eea
In the following the quantities in three dimensions 
are denoted by the suffix ``$(3)$'' and the quantities 
without the suffix are those in two dimensions unless 
we mention.
In the parametrization (\ref{3dmtrc}),  
curvature  has the following form 
\be
\label{cvsclr}
R^{(3)}=R+\Box \phi - \partial_\alpha\phi
\partial^\alpha\phi - {1 \over 4}
\e^{-2\phi}F_{\alpha\beta}F^{\alpha\beta}\ .
\ee
Here $F_{\alpha\beta}$ is the field strength:
$F_{\alpha\beta}=\partial_\alpha A_\beta - \partial_\beta A_\alpha$.
Then the action $S$ of the gravity with  cosmological term 
and with  $N$ free scalars $f_i$ 
($i=1,\cdots,N$) in three dimensions is 
reduced as
\bea
\label{redS}
S&=&{1 \over 4\pi^2} 
\int d^3x \sqrt{g^{(3)}}\left\{{1 \over 8G}\left( R^{(3)} + \Lambda \right)
+ {1 \over 2}\sum_{i=1}^N \partial_\mu f_i \partial^\mu f^i \right\}\nn
&\sim&{1 \over 2\pi} \int d^2x \sqrt{g}\e^{-\phi}\left\{
{1 \over 8G}\left(R - {1 \over 4} \e^{-2\phi}F_{\alpha\beta}F^{\alpha\beta}
+\Lambda \right)+ {1 \over 2}\sum_{i=1}^N \partial_\alpha f_i 
\partial^\alpha f^i \right\}\ .
\eea
Note that the contribution from 
the second and third terms in (\ref{cvsclr}) becomes 
the total derivative which maybe neglected.
 One BH solution for the classical action is given by \cite{BTZ}
\bea
\label{BTZclass}
&& ds^2=-\e^{2\rho_{cl}}dt^2 + \e^{-2\rho_{cl}}dx^2 \ ,\ \ 
\e^{2\rho_{cl}}=-M+{x^2 \over l^2} + {J^2 \over 4x^2} \nn
&& \e^{-\phi}=\e^{-\phi_{cl}}=\e^{-\phi_0}x
\eea
Here ${1 \over l^2}={\Lambda \over 2}$. $M$ is the parameter related 
to the black hole mass and $J$ is that of the angular momentum in 
 3D model or the electromagnetic charge in the 2D  
 model.
The extremal limit is given by $J^2=l^2 M^2$. 
In order to consider this limit, we change the coordinates as follows
\be
\label{chgcoord}
x^2={l^2 M \left( 1 + \epsilon \tanh r \right) \over 2}\ ,\ \ \ 
t={l \over \epsilon\sqrt{2}}\tau\ .
\ee
Then $r\rightarrow +\infty$ corresponds to outer horizon and 
$r\rightarrow -\infty$ corresponds to inner horizon.
 Taking the limit of $\epsilon\rightarrow 0$, 
we obtain
\be
\label{extsol}
ds^2={l^2 M \over 4\cosh^2 r}\left(d\tau^2 - dr^2\right)\ ,\ \ \ 
\e^{-\phi}={l\e^{-\phi_0}}{\sqrt{M \over 2}}\ .
\ee
Note that $\phi$ becomes a constant in this limit.

\ 

\noindent
3. We will discuss now the quantum corrections induced by
$N$ free conformally invariant dilaton coupled 
scalars $f_i$:\footnote{Quantum corrections due to matter near 
BTZ black hole have been also studied in \cite{BVZ}.}
\be
\label{sS}
S^f=-{1 \over 2}\int d^2x \sqrt{-g}\e^{-\phi}
 \sum_{i=1}^N(\nabla f_i)^2\ .
\ee 
This action represents reduction of 3D minimal scalars action.
We present now the two-dimensional metric as 
$ds^2=\e^{2\sigma}\tilde g_{\mu\nu}d x^\mu d x^\nu$, 
where  $\tilde g_{\mu\nu}$ is the flat metric.
Then the total quantum action coming from quantum scalars 
is given as (we work in large $N$ approximation what 
justifies the neglecting of proper quantum 
gravitational corrections):
\be
\label{Gamma}
\Gamma = W + \Gamma[1, \tilde g_{\mu\nu}]
\ee
where $W$ is conformal anomaly induced effective action \cite{BH,NO} 
(actually, scale-dependent part) 
and $\Gamma[1, \tilde g_{\mu\nu}]$ is one-loop 
conformally invariant effective 
action for scalars (\ref{sS}) calculated on flat 2D metric. 

Following \cite{NO} ( using conformal anomaly of dilaton coupled scalar 
\cite{BH,NO,A}) we obtain
\be
\label{qc2}
W=-{1 \over 2}\int d^2x \sqrt{-g} \left[ 
{N \over 48\pi}R{1 \over \Delta}R  -{N \over 16\pi} \nabla^\lambda \phi
\nabla_\lambda \phi {1 \over \Delta}R +{N \over 8\pi}\phi R \right] \ .
\ee
We are left with the calculation of the conformally 
invariant part of the effective action $\Gamma[1, \tilde g_{\mu\nu}]$. 
This term has to be calculated on the flat space, 
after that the general covariance should be restored. 
It is impossible to do such calculation in closed form. 
We apply Schwinger-De Witt technique in order to get this effective action 
as local curvature expansion.

Using the results of ref.\cite{NO}, one can obtain:
\be
\label{Gamma1}
\Gamma[1, \tilde g_{\mu\nu}]={N \over 16\pi}\int d^2x 
\sqrt{-\tilde g}(-\tilde \nabla_\mu \phi 
\tilde \nabla^\mu \phi ) \ln \mu^2 + \cdots 
\ee
Here $\mu$ is dimensional parameter.\footnote{
Note that more exactly, one should write 
$\ln {L^2 \over \mu^2}$ where $L^2$ is covariant cut-off parameter.}
 In general covariant form we have:
\be
\label{Gamma1b}
\Gamma[1, g_{\mu\nu}]=-{N \over 16\pi}\int d^2x  
\sqrt{- g}\nabla_\mu \phi \nabla^\mu \phi \ln \mu^2 + \cdots \ .
\ee
This action is conformally invariant as it should be in 
accordance with Eq.(\ref{Gamma}).

\ 

\noindent
4. In the study of quantum corrected BH we 
work in the conformal gauge: 
$g_{\pm\mp}=-{1 \over 2}\e^{2\rho}\ ,\ \ g_{\pm\pm}=0$. 
We are going to study (in)stability of black holes in the 
same way as in four dimensions \cite{NO2}.

The equations of motion with account of quantum 
corrections are given by
\bea
\label{eqnppBTZ}
0&=&{1 \over 8G}\e^{-\phi}\left(-\partial_\pm \rho \partial_\pm\phi 
- {1 \over 2} \partial_\pm^2\phi + {1 \over 2} (\partial_\pm\phi)^2\right) 
+{N \over 12}\left( \partial_\pm^2 \rho 
- \partial_\pm\rho \partial_\pm\rho \right) \nn
&& +{N \over 8} \left\{\left( \partial_\pm \phi \partial_\pm\phi \right)
\rho+{1 \over 2}{\partial_\pm \over \partial_\mp}
\left( \partial_\pm\phi \partial_\mp\phi \right)\right\} \nn
&& +{N \over 8}\left\{ 2 \partial_\pm \rho \partial_\pm \phi 
-\partial_\pm^2 \phi \right\} + t^\pm(x^\pm) 
+{N \over 8} \left( \partial_\pm \phi\partial_\pm\phi \right)\ln \mu^2 \\
\label{reqBTZ}
0&=&{1 \over 8G}\e^{-\phi}\left(4\partial_+ \partial_- \phi 
- 4 \partial_+\phi\partial_-\phi 
- \Lambda \e^{2\rho} + 2\e^{-2\phi-2\rho}(F_{+-})^2\right) \nn
&& -{N \over 12}\partial_+\partial_- \rho
-{N \over 16}\partial_+ \phi \partial_- \phi 
+{N \over 8}\partial_+\partial_-\phi \\
\label{eqtpBTZ}
0&=& {1 \over 8G}\e^{-\phi}\left(4\partial_+ \partial_- \rho
+ {\Lambda \over 2} \e^{2\rho}
+ 3\e^{-2\phi-2\rho}(F_{+-})^2\right)  \nn
&& -{N \over 16}\partial_+(\rho \partial_-\phi)
-{N \over 16}\partial_-(\rho \partial_+\phi)
-{N \over 8}\partial_+\partial_-\rho  
-{N \over 4} \partial_+ \partial_- \phi \ln \mu^2\\
\label{eqABTZ}
0&=&\partial_\pm(\e^{-3\phi-2\rho}F_{+-})\ .
\eea
Eq.(\ref{eqABTZ}) can be integrated to give
\be
\label{eqA2BTZ}
\e^{-3\phi-2\rho}F_{+-}=B\ (\mbox{constant})\ .
\ee
We now assume $\phi$ is constant as in the classical extremal solution 
(\ref{extsol})  even when we include the quantum correction:
$\phi=\phi_0\ (\mbox{constant})$. Then we obtain
\be
\label{constRBTZ}
B^2={\Lambda \e^{-4\phi_0} \over 2 } \ ,\ \ 
R=R_0
\equiv -{4\Lambda \e^{-\phi_0} \over 
\e^{-\phi_0}-{GN \over 4} }\ .
\ee
Quantum corrected extremal (static) solution 
corresponding to the classical one (\ref{extsol}) can be 
written as follows:
\be
\label{rho0}
\e^{2\rho}=\e^{2\rho_0}\equiv {2C \over R_0}
\cdot {1 \over \cosh^2 \left(r\sqrt{C} \right)}\ .
\ee
Here $r={x^+ - x^- \over 2}$ and $C$ is an integration constant.

In order to investigate the (in)stability in the above solution, 
we consider  
the perturbation around the extremal solution:
\be
\label{pert}
\rho=\rho_0 + \epsilon R\ ,\ \ \ \phi=\phi_0 +\epsilon S\ .
\ee
 Following Bousso and Hawking's argument\cite{BH2}, we neglect 
the second term in (\ref{qc2}).
Then (\ref{reqBTZ}) and (\ref{eqtpBTZ}) (and (\ref{eqA2BTZ})) 
  lead to the following equations:
\bea
\label{reqpertBTZ}
0&=&{1 \over 8G}\e^{-\phi_0}\left\{4\partial_+\partial_-S 
- \Lambda \e^{2\rho_0} (2R-S) + 2B^2 \e^{2\rho_0+4\phi_0}(2R+3S)\right\} \nn
&& - {N \over 12}\partial_+\partial_- R+ {N \over 8}\partial_+\partial_-S \\
\label{eqtppertBTZ}
0&=&{1 \over 8G}\e^{-\phi_0}\left\{4\partial_+\partial_-R
- 4\partial_+\partial_- \rho_0 S
+ {\Lambda \over 2} \e^{2\rho_0} (2R-S)\right.\nn
&& \left. + 3B^2 \e^{2\rho_0+4\phi_0}
(2R+3S)\right\} - {N \over 8}\partial_+\partial_- R 
-{N \over 4}\ln \mu^2 \partial_+\partial_- S \ .
\eea
We now assume $R$ and $S$ have the following form
\be
\label{RS}
R(t,r)=P\cosh\left(t\alpha\sqrt{C}\right)\cosh^\alpha r\sqrt{C} \ ,\ \ 
S(t,r)=Q\cosh\left(t\alpha\sqrt{C}\right)\cosh^\alpha r\sqrt{C} \ .
\ee
Then (\ref{reqpertBTZ}) and (\ref{eqtppertBTZ}) become 
algebraic equations and the condition that there is  
non-trivial solution for $Q$ and $P$ is given by
\bea
\label{detBTZ}
F^{BTZ}(A)&\equiv&\e^{-2\phi_0}\left({2R_0 \over C}A + 4\Lambda\right)^2 
- \left({GNR_0 \over C}A\right)^2 \nn
&& + {GNR_0 \over 3C}A\left\{\e^{-\phi_0}\left(-{R_0 \over 2}
+ 4\Lambda\right) - GN (\ln\mu^2) {R_0 \over C}A\right\}=0\ .
\eea 
Here $A={\alpha(\alpha -1 ) C \over 4}$. $\e^{-\phi}$ is identified with 
the radius  in 3D model, so horizon is given by the condition
\be
\label{horizon}
\nabla\sigma\cdot\nabla\sigma=0\ .
\ee
Substituting (\ref{RS}) into (\ref{horizon}), we find the horizon is given 
by $r=\pm\alpha t$. Therefore on the horizon, we obtain
\be
\label{Shrzn}
S(t,r(t))=Q\cosh^{1+\alpha} t\alpha\sqrt{C} \ .
\ee
This tells that the system is unstable if there is a solution $0>\alpha >-1$, 
i.e., $0<A<{C \over 2}$. On the other hand, the perturbation becomes stable 
if there is a solution where $\alpha<-1$, i.e., $A>{C \over 2}$. The 
radius of the horizon $r_h$ is given by
\be
\label{hrds}
r_h=\e^\sigma =\e^{-\left(\phi_0 + \epsilon S(t,r(t))\right)}\ .
\ee
Let the initial perturbation is negative $Q<0$. Then the radius shrinks 
monotonically, i.e., the black hole evaporates 
(due to quantum effects) in case of $0>\alpha>-1$.
On the other hand, the radius increases in time and approaches to the 
extremal limit asymptotically: 
$S(t,r(t))\rightarrow Q\e^{(1+\alpha) t |\alpha|\sqrt{C}}$
in case of $\alpha<-1$. The latter case corresponds to the quantum 
anti-evaporation
of Nariai black hole observed by Bousso and Hawking \cite{BH2} in
case of 4D BH with minimal scalars (for the case of conformal matter 
induced anti-evaporation, see\cite{NO2}).
In the classical limit $\e^{-2\phi_0}\gg N$, $A={C \over 2}$, 
what tells that there does not occur any kind of the radiation 
in the solution.  

Eq.(\ref{detBTZ}) can be solved with respect to $A$,
\bea
\label{AC}
{A \over C}&=&\left[ -4096 - 2176 g + 48 g^2 
\pm \left\{ 18939904 g + 4321280 g^2 \right.\right.\nn
&& - 208832 g^3 + 2304 g^4 + \left(786432 g - 49152 g^2 \right. \nn
&& \left.\left.\left. + 768 g^3 \right)\ln\mu^2
\right\}^{1 \over 2}\right] 
 \times \left\{2\left( -4096 + 16 g^2 + 192 g^2 
\ln\mu^2 \right)\right\}^{-1}\nn
&\sim&{1 \over 2} \pm {3 \sqrt{2} \over 4}g^{1 \over 2} 
+ {\cal O}(g) 
\eea
Here $g \equiv 8GN\e^{\phi_0}$.
Note that there are two solutions near the classical limit $g\rightarrow 0$, 
which correspond to stable and instable modes, respectively. It might be 
surprizing that there is an instable mode since the extremal solution is 
usually believed to be stable.
The global behavior of ${A \over C}$ is given in Figures.
In Fig.1, the vertical line corresponds to ${A \over C}$ and 
the horizontal one to $g$ when $\ln\mu^2=1$. There is a 
singularity when $g\sim 0.4376$. The singularity occurs when the denominator 
in (\ref{AC}) vanishes. In the range from $g=0$ 
to near the singularity, there coexist the stable and instable modes. 
In Fig.2a and 2b, we present the 
 global behavior of two modes as a 
function of $g\equiv 8GN\e^{\phi_0}$ and $\ln\mu^2$. 
Fig.2a corresponds to the stable mode and 2b to the instable one.

\ 

\noindent
5.  In the recent study by Bousso\cite{B} it has been proposed  
the possibility that de Sitter space proliferates.
 Consider Schwarzschild de Sitter (SdS) BH in an extremal 
 regime (Nariai BH). Topology of Nariai 
space is $S^1\times S^2$, therefore the solution can describe 
a handle. If there is a perturbation, multiple 
pair of the cosmological and black hole horizons are formed. After that 
 BHs evaporate and BHs horizons tend to zero.
Therefore the handle is separated into several pieces, which are copies 
of the de Sitter spaces. The above analysis has been supported by 
the study of higher modes perturbations in the spectrum of Nariai BH. 

The metric (\ref{extsol}) in the extremal limit or its quantum analogue 
(\ref{rho0}) has the same spacetime structure as the Nariai solution except  
its signature. Therefore there is a possibility of the multiple production 
of universes from the extremal soltion in 3D BTZ BH also.

Let us define new coordinate $\theta$ 
\be
\label{theta}
\sin\theta = \tanh \left(r\sqrt{C}\right)
\ee
Then metric corresponding to (\ref{rho0}) becomes
\be
\label{thetametric}
ds^2={2C \over R_0}\left(-dt^2 + {1 \over C}d\theta^2\right)\ .
\ee
Since $\theta$ has the period $2\pi$, the topology of the space with 
the metric (\ref{thetametric}) can be $S_1$. Therefore the topology 
of the space in the original BTZ model is the torus of $S_1\times S_1$.
Note that Eq.(\ref{theta}) tells there is one to two correspondence 
between $\theta$ and $r$.

The functions $R$ and $S$ in (\ref{RS}) correspond to 
 the eigenfunction with the 
eigenvalue $A$ (\ref{detBTZ}) of the operator
\be
\label{Delta}
\Delta\equiv C\cosh^2\left(r\sqrt{C}\right)\partial_+\partial_- 
\ee
which can be regarded as the Laplacian on two-dimensional 
Lorentzian hyperboloid, i.e., the Casimir operator of $SL(2,R)$. The 
operators of $SL(2,R)$ are given by
\be
\label{rl}
L_0={1 \over \sqrt{C}}{\partial \over \partial t}\ ,\ \ 
L_\pm={1 \over \sqrt{C}}\e^{\pm t\sqrt{C}}\left(\sinh\left(r\sqrt{C}\right)
{\partial \over \partial t} \pm \cosh\left(r\sqrt{C}\right){\partial \over \partial r}\right)\ .
\ee
The operators $L_0$ and $L_\pm$ form the algebra $SL(2,R)$
\be
\label{sl2alg}
[L_+,L_-]=2L_0\ ,\ \ [L_0, L_\pm]=\pm L_\pm\ .
\ee
The eigenfunction of $\Delta$ in (\ref{RS}) is the sum of highest and 
lowest weight representation $\e^{\pm t\alpha\sqrt{C}}
\cosh^\alpha r\sqrt{C}$. We now consider the following eigenfunction
\bea
\label{S2}
S(t,r)&=&{Q \over 4} \left(L_+\e^{ t\alpha\sqrt{C}}\cosh^\alpha r\sqrt{C}
+ L_-\e^{- t\alpha\sqrt{C}}\cosh^\alpha r\sqrt{C}\right) \nn
&=&Q\sinh\left(t(\alpha+1)\sqrt{C}\right)\cosh^\alpha r\sqrt{C}
\sinh r\sqrt{C}\ .
\eea
Since the evaporation of the black hole in \cite{B} corresponds 
to the instable mode, we restrict $\alpha$ to be $0>\alpha>-1$.
Then  using the condition of horizon (\ref{horizon}) we may estimate 
 the asymptotic behavior of $S$ 
\bea
\label{Sasym}
S(t,r(t))&\stackrel{t\rightarrow 0}{\rightarrow}&\pm (\alpha +1)CQ 
t^2 \nn
&\stackrel{t\rightarrow +\infty}{\rightarrow}&\pm{Q \over 2^{\alpha+2}}
\left({1-\alpha \over 1+\alpha }\right)^{\alpha + 1 \over 2}
\e^{t(\alpha + 1)(\alpha + 2)\sqrt{C}}
\eea
Since the radius of the horizon is given by (\ref{hrds}), 
the horizon corresponding to $+$ sign in (\ref{Sasym}) grows up to 
infinity and that to $-$ sign tends to zero when 
$t\rightarrow +\infty$ although the perturbation 
does not work when $S$ is large. The horizon of $+$ sign corresponds to 
outer horizon and that of minus sign to inner horizon. Since there is 
one to two correspondence between the radial coordinates $r$ and $\theta$, 
there are two outer horizons and two inner ones. When we regard the 
model as 3D model, the handle with the 
topology of $S_1\times S_1$ is separated to two pieces when the two 
inner horizons tend to zero. Of course, this is 
qualitative discussion as perturbation should be small. 
 Above result can be generalized if 
we use higher order eigenfunctions as perturbation
\be
\label{Sn}
S(t,r)={Q \over 4} \left((L_+)^n \e^{ t\alpha\sqrt{C}}\cosh^\alpha r\sqrt{C}
+ (L_-)^n\e^{- t\alpha\sqrt{C}}\cosh^\alpha r\sqrt{C}\right)\ .
\ee
Since the condition of the horizon (\ref{horizon}) gives the $2n$-th 
algebraic equation with respect to $\tanh r\sqrt{C}$, 
there can appear $2n$ outer horizons and $2n$ inner horizons and 
the spacetime could be disintegrated into $2n$ pieces when the inner 
horizons tend to zero in accordance with proliferation 
proposal \cite{B}.
Note that horizon equation for the higher mode is $2n$-th order 
polynomial with respect to $\tanh r\sqrt{C}$, which is the 
reason why it is expected that there can be $2\times 2n$ 
horizons. However, it is difficult to confirm that all solutions 
are real solutions.

\ 

\noindent
6. In summary, we investigated quantum evolution of BTZ BH in large $N$ 
approximation. As we show the presence of instable modes in quantum 
spectrum manifests in anti-evaporation or possible topological 
proliferation. It is clear that in order to understand if above 
effects may really happen in early Universe one should select 
boundary conditions corresponding to BTZ  BH formation and analyse 
above processes subject to such boundary conditions. Depending on 
the choice of boundary conditions they maybe compatible with 
anti-evaporation as it happens for Nariai anti-evaporating BH \cite{NO2}. 

Our study is quite general and we may expect that similar results 
could be true for higher dimensional BHs described as product of BTZ 
sector and (anti)sphere. This is already confirmed \cite{NO3}
for 2D charged BH 
\cite{MNY} (which is dual to 5D BH \cite{Teo}) in the similar large $N$ approach 
used in the calculation of one-loop effective action for dilaton 
coupled scalars.

\ 

\noindent
{\bf Acknoweledgments} We thank R. Bousso for useful discussion.

\ 

\noindent
{\Large\bf Figure Captions}

\ 

\noindent
Fig.1 ${A \over C}$ (vertical line) versus 
$g\equiv 8GN\e^{\phi_0}$ (horizontal line)
when $\ln\mu^2=1$.

\ 

\noindent
Fig.2a,b Two branches of ${A \over C}$ (vertical line) versus 
$g\equiv 8GN\e^{\phi_0}$ in $[0,1]$ and $\ln\mu^2$ in
$[0,3]$. Fig.2a corresponds to the stable mode and 2b to the 
instable one.

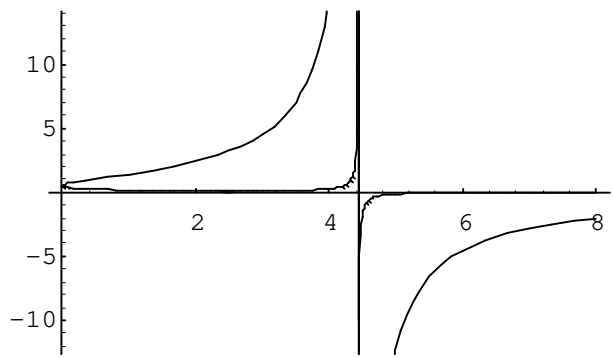
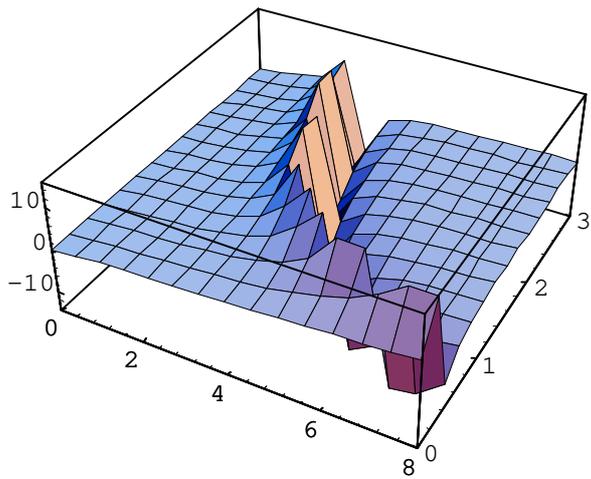
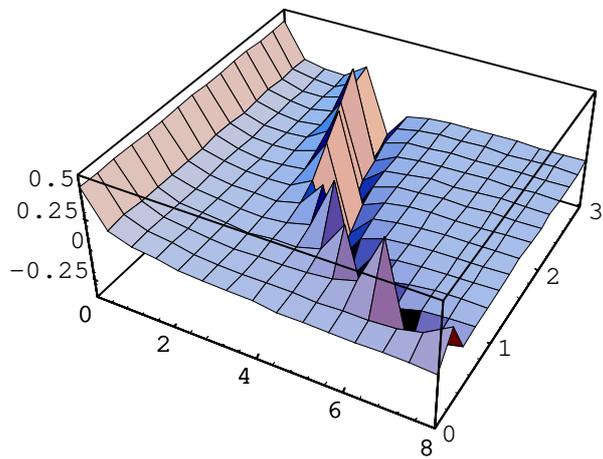
\end{document}